\documentclass[9pt,twocolumn,twoside]{pnas-new_JHEP}

\makeatletter
\newcommand{\manuallabel}[2]{\def\@currentlabel{#2}\label{#1}}
\makeatother

\setboolean{displaywatermark}{false}
\templatetype{pnasresearcharticle} 
\usepackage{amsfonts}
\usepackage{upgreek}
\usepackage{slashed}
\usepackage{amsmath,amssymb,bm,bbm} 
\usepackage{latexsym}
\usepackage[export]{adjustbox}
\usepackage[colorlinks=true]{hyperref}  
\usepackage[toc,page]{appendix}
\usepackage{dsfont}
\usepackage{pdfpages}
\usepackage{graphicx}
\usepackage{subfig}
\usepackage{siunitx}
\pdfoutput=1

\definecolor{wrongultramarine}{rgb}{1,0.5,0}

\newcommand{\appref}[1]{Sec.~\ref{#1} of the SI}

\title{Quantum oscillations in the hole-doped cuprates and the confinement of spinons}

\author[a]{Pietro M. Bonetti\textsuperscript{1}}
\author[a]{Maine Christos}
\author[a]{Subir Sachdev}

\affil[a]{Department of Physics, Harvard University, Cambridge MA-02138, USA}

\leadauthor{Bonetti} 

\significancestatement{
The most robust realization of high temperature superconductivity at ambient pressure is provided by the copper-oxide based `cuprate' materials. In the hole-doped case, the parent state above the superconducting critical temperature is an unconventional metal with a `pseudogap', and this has been the focus of much experimental and theoretical study. In high magnetic fields, $B$, and low temperatures, the pseudogap metal exhibits quantum oscillations: numerous physical properties oscillate with characteristic periods as a function of $1/B$. We propose a long-sought explanation for these oscillations, using spinon excitations of an underlying quantum spin liquid which are fractions of a single electron. Our results open a route to understanding the quantum spin liquid mechanism for high temperature superconductivity. 
}

\correspondingauthor{\textsuperscript{1}To whom correspondence should be addressed. E-mails: pbonetti@fas.harvard.edu and sachdev@g.harvard.edu}

\keywords{high temperature superconductors $|$ doped spin liquids $|$ quantum oscillations} 

\begin{abstract}
 A long standing problem in the study of the under-hole-doped cuprates has been the description of the Fermi surfaces underlying the high magnetic field quantum oscillations, and their connection to the higher temperature pseudogap metal. Harrison and Sebastian (\href{https://journals.aps.org/prl/abstract/10.1103/PhysRevLett.106.226402}{Phys. Rev. Lett. {\bf 106}, 226402 (2011)}) proposed that the pseudogap `Fermi arcs' are reconstructed into an electron pocket by field-induced charge density wave order. But computations on such a model (Zhang and Mei, \href{https://iopscience.iop.org/article/10.1209/0295-5075/114/47008}{Europhys. Lett. {\bf 114}, 47008 (2016)}) show an unobserved additional oscillation frequency from a Fermi surface arising from the backsides of the hole pockets completing the Fermi arcs. We describe a transition from a fractionalized Fermi liquid (FL*) model of the pseudogap metal, to a metal with bi-directional charge density wave order without fractionalization. We show that the confinement of the fermionic spinon excitations of the FL* across this transition can eliminate the unobserved oscillation frequency. 
\end{abstract}

\dates{This manuscript was compiled on \today}

\begin{document}

\maketitle
\ifthenelse{\boolean{shortarticle}}{\ifthenelse{\boolean{singlecolumn}}{\abscontentformatted}{\abscontent}}{}
\noindent
\href{https://arxiv.org/abs/2405.08817}{arXiv:2405.08817} 



\dropcap{D}ecades of careful experimental study of the hole-doped copper oxide based high temperature superconductors have revealed a remarkable evolution of their low energy fermionic excitations in the underdoped regime. 
In the higher temperature pseudogap metal, photoemission \cite{Johnson11,He2011,chen2019incoherent} and scanning tunneling microscopy \cite{Hoffman14,Davis14} display `Fermi arcs' which have been described by excitations of a hole pocket model \cite{SS94,WenLee96,WenLee98,Stanescu_2006,Berthod_2006,Yang_2006,Kaul07,KaulKim07,Sakai_2009,Qi_2010,Mei_2012,Robinson_2019,Ancilla1,Mascot22,Skolimowski_2022,Fabrizio23,Giorgio23,Bonetti2022,Weng23}. In contrast, at low temperatures and high magnetic fields, quantum oscillations \cite{Louis07,Yelland2008,Sebastian10,Boebinger11,Sebastian11,Barisic2013,Greven16,SuchitraCyril,Proust_2024} are consistent with the excitations of an electron pocket model \cite{Suchitra-electron,Atkinson15,Allais14,Zhang_2016,Proust_2024}. A theoretical understanding of the evolution between these distinct Fermi surfaces at high and low magnetic fields remains a central open problem in the study of the cuprates. 
A crucial ingredient in the evolution from Fermi arcs to electron pockets is the field-induced charge density wave (CDW) order that is experimentally known to set in at high fields \cite{Greven14,Greven17,Proust_review}. 

In the present paper we argue that charge neutral, spin-1/2, fermionic spinons are another important and necessary ingredient to reproduce the experimental data. We present a model that has spinon excitations with massless Dirac dispersion in the
higher temperature pseudogap metal, and these spinons confine across the transition involving the onset of CDW order at low temperatures. Our model displays key features of the observations which have been difficult to reconcile so far:\\
({\it i\/}) It describes the `Fermi arc' spectra and other features of the higher temperature pseudogap metal, as was already discussed in Ref.~\cite{Mascot22}.\\
({\it ii\/}) The spectrum of quantum oscillations in the low temperature, high field CDW state in our model is consistent with only a single electron pocket, as has been argued from the experimental observations \cite{Sebastian11,Greven16}.\\
({\it iii \/}) It can account for the excess linear-in-temperature specific heat observed in the high field CDW state in HgBa$_{2}$CuO$_{4+\delta}$ \cite{Girod20}.

\begin{figure}[ht]
    \includegraphics[width=3.5in]{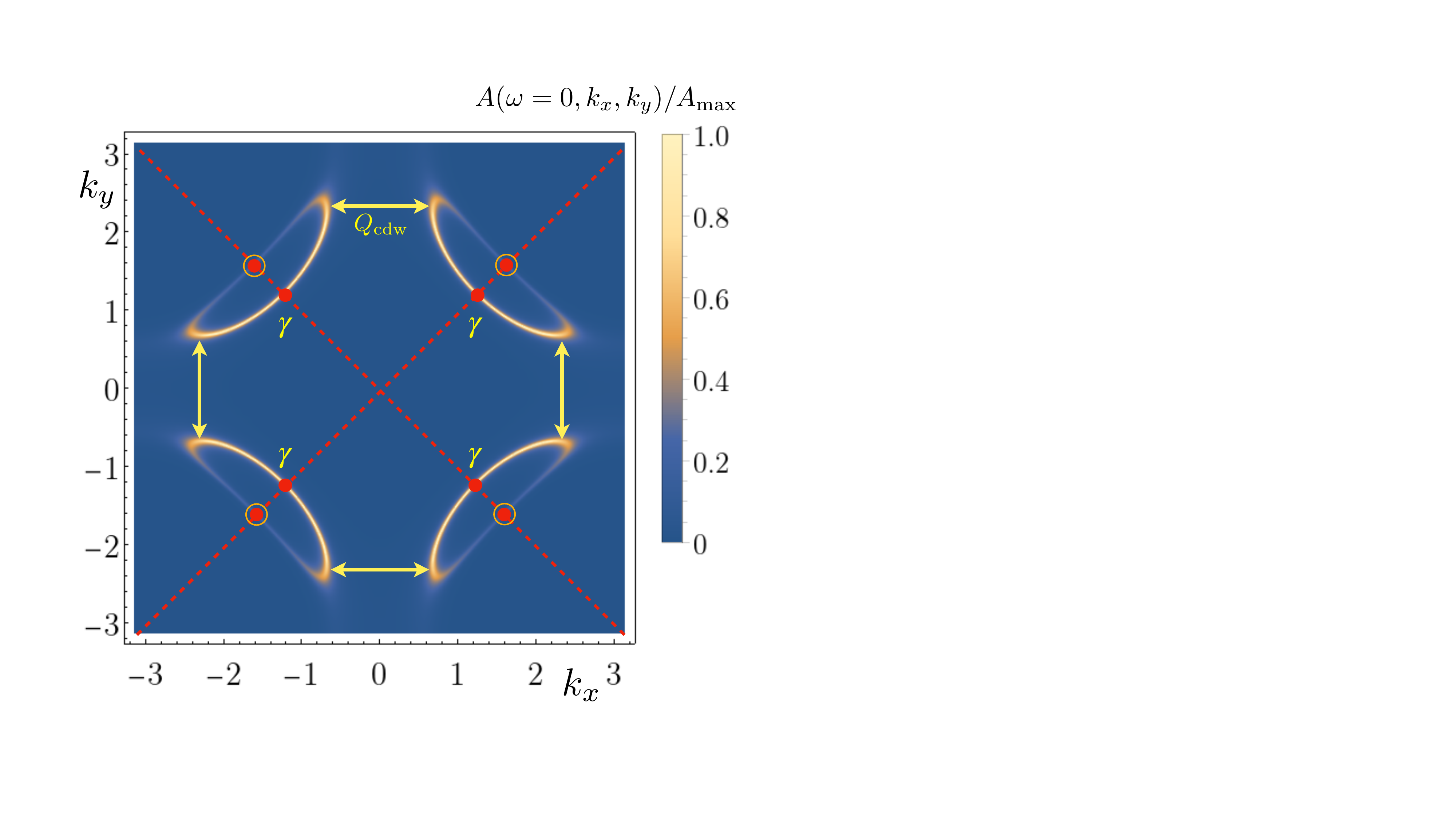}
    \caption{Color density plot of the electron spectral density $A(\omega, k_x, k_y)$ (frequency $\omega$ and wavevector $(k_x, k_y)$) computation in the FL* pseudogap metal \cite{Mascot22} without CDW order or $d$-wave superconductivity, showing the four $\gamma$ hole pockets (as denoted in Fig.~\ref{fig:3}a) which enclose the non-Luttinger area corresponding to hole density $p$. The `front sides' of the $\gamma$ hole pockets are the bright Fermi arcs, while the faint `backsides' have not been detected in photoemission. The dashed lines indicate the zeros of the pairing amplitude in the $d$-wave superconductor. The eight filled circles indicate the nodal points of Bogoliubov quasiparticles obtained when $d$-wave pairing is imposed on the hole pockets. The four open circles indicate the positions of the nodal fermionic spinons of the critical spin liquid \cite{Affleck1988}: these fermionic nodal spinons annihilate the four nodal Bogoliubov quasiparticles on the backsides in a $d$-wave superconductor in which the spin liquid confines, leaving only four nodal Bogoliubov quasiparticles on the front sides. The ordering wavevectors of the CDW are $Q_{\rm cdw}$; connecting the front sides of the hole pockets yields the electron pocket model of quantum oscillations \cite{Suchitra-electron}. Fermi surfaces with CDW order present are not shown in this figure.}
    \label{fig:1}
\end{figure}

A key to the resolution of the experimental observations has been an understanding of the role of the `backside' of the hole pockets (the $\gamma$ pockets of Figs.~\ref{fig:1} and \ref{fig:3}) presumed to complete Fermi arcs of the pseudogap metal---see Fig.~\ref{fig:1}. 
This `backside problem' shows up in a number of experimental observables, as we now describe. \\

\noindent\textbf{The backside problems.}\\
\noindent
(A) {\it Photoemission in the pseudogap metal.}\\
In almost all metals, the low energy fermionic excitations are charge $\pm e$, spin-1/2 quasiparticles on a Fermi surface enclosing a volume dictated by the Luttinger relation. But in the higher temperature pseudogap, photoemission observations reveal a `Fermi arc' spectrum along the Brillouin zone diagonals. This is most commonly modeled by hole pockets with Fermi surfaces that enclose a non-Luttinger area $p$ at hole density $p$ \cite{SS94,WenLee96,WenLee98,Stanescu_2006,Berthod_2006,Yang_2006,Kaul07,KaulKim07,Sakai_2009,Qi_2010,Mei_2012,Robinson_2019,Ancilla1,Mascot22,Skolimowski_2022,Bonetti2022,Fabrizio23,Giorgio23,Weng23} (the Luttinger area at this doping is $1+p$). As shown in Fig.~\ref{fig:1}, such $\gamma$ hole pockets have an approximately elliptical shape, and the observed Fermi arcs are the `front side' of the ellipse. The `backside' has not been observed in photoemission or tunnelling experiments; it has a low spectral intensity in theoretical models, and is presumably further suppressed from the mean-field computation in Fig.~\ref{fig:1} by quantum fluctuations associated with spinons \cite{Mascot22} and impurities.\\

\noindent
(B) {\it Extra frequency in quantum oscillations}\\
The backside problem is also manifested in the quantum oscillations at low temperatures and high fields. A computation \cite{Zhang_2016} of the fermionic spectrum in a model of hole pockets in the presence of field-induced CDW order yields the observed $\alpha$ electron pockets \cite{Zhang_2016} (see Fig.~\ref{fig:3}) by connecting together the hole pocket front sides by the CDW ordering wavevector, as was argued by Harrison and Sebastian \cite{Suchitra-electron} (Fig.~\ref{fig:1}). However, in the presence of CDW order, in addition to the $\alpha$ pockets, the backsides of the hole pockets also lead to a second electron-like $\beta$ Fermi surface \cite{Zhang_2016} (Fig.~\ref{fig:3}) which has not been observed.\\

\noindent
(C) {\it Eight nodal points in the $d$-wave superconductor}\\
A third manifestation of the backside problems arises when we consider the transition from the pseudogap metal to the superconducting state in the absence of CDW order. Imposing a $d$-wave pairing gap upon the $\gamma$ hole pocket model of the pseudogap leads to {\it eight\/} nodal points of Bogoliubov quasiparticles in the $d$-wave superconductor (sketched in Fig.~\ref{fig:1}) because, without long-range antiferromagnetic order, the first Brillouin zone remains that of the underlying square lattice. However, there is much experimental evidence for only {\it four\/} nodes in the superconducting state \cite{Chiao00}, and such thermal conductivity and specific heat measurements preclude a scenario where there are additional nodes with a low spectral intensity in the superconducting state.

A resolution of the backside problem (C) in the superconducting state was proposed in Refs.~\cite{ChatterjeeSS16,Christos23} using charge neutral spin-1/2 fermionic spin excitations of the pseudogap. Such fractionalized excitations are required to be present whenever there are Fermi surfaces at zero temperature which do not enclose the Luttinger volume \cite{MO00,FLS1,FLS2,Arun04}, leading to a `fractionalized Fermi liquid' (FL*). Christos {\it et al.\/} \cite{PNAS_pseudo} proposed that the fermionic spinon excitations are those of the multi- or pseudo-critical spin liquid \cite{Zhou:2023qfi,Meng23,Chester:2023njo,Sandvik24} of the N\'eel-valence bond solid transition \cite{NRSS89}, whose mean-field realization \cite{DQCP3} is the $\pi$-flux phase with four nodal points \cite{Affleck1988} (see Fig.~\ref{fig:1}; this spin liquid also exhibits $d$-wave superconductivity upon doping \cite{Jiang21,Jiang23}). In the low energy theory of Ref.~\cite{PNAS_pseudo}, the Higgs condensation of the boson which couples the physical electrons to the fractionalized spinons drives transitions to various symmetry-breaking phases, including charge order and $d$-wave superconductivity. The spin liquid must confine in a conventional superconductor: in the confined superconducting phase, the four spinon nodal points of the spin liquid are allowed to mix and annihilate with the four Bogoliubov quasiparticle nodal points on the backsides of the $\gamma$ hole pockets. This leaves just the required four Bogoliubov nodal points associated with the front sides in the confining $d$-wave superconductor. 

In the present paper, we exploit the 
fermionic spinons of the $\pi$-flux spin liquid to address the backside problem (B) of the low temperature, high field quantum oscillations. In analogy with the case where the ordered phase is a $d$-wave superconductor, we assume that the spin liquid is confined across the transition from the FL* pseudogap metal to the CDW state \cite{PNAS_pseudo,Patel:2016efz} (where neither phase has antiferromagnetic order). The Higgs condensation of the boson which couples the physical electrons and spin liquid allows the fermionic spinons to mix with the excitations on the Fermi surfaces in the CDW state. We find that this mixing can eliminate the unobserved $\beta$ quantum oscillations, leaving prominent quantum oscillations only from the $\alpha$ electron pocket of Harrison and Sebastian (Fig.~\ref{fig:4}b). 
%
A Fermi surface consisting of a single electron pocket can also be obtained in a state with coexisting CDW and (incommensurate) antiferromagnetic order~\cite{Eberlein2016,Proust_2024}.
The confinement of spinons across the FL*$\rightarrow$CDW transition also leaves small $\delta$ pockets (Fig.~\ref{fig:3}c) which we propose as the origin of the excess linear-in-temperature specific heat \cite{Girod20}.

We will begin in Section~\ref{sec:ancilla} by introducing the `ancilla' approach \cite{Ancilla1} which allows for a simple, microscopic description of the FL* phase, and its confinement transitions, in a single-band Hubbard model. We note, however, that the ancilla approach is not essential for our results, and similar results can also be obtained in more phenomenological approaches which do not introduce ancilla degrees of freedom \cite{ChatterjeeSS16}. We describe the Fermi surfaces in the phases of the ancilla model in Section~\ref{sec:fermi}, and their quantum oscillations in a magnetic field in Section~\ref{sec:qo}. These results are connected to experimental observations in Section~\ref{sec:conc}.

\section{The ancilla model}
\label{sec:ancilla}

\begin{figure}[t]
    \includegraphics[width=3.5in]{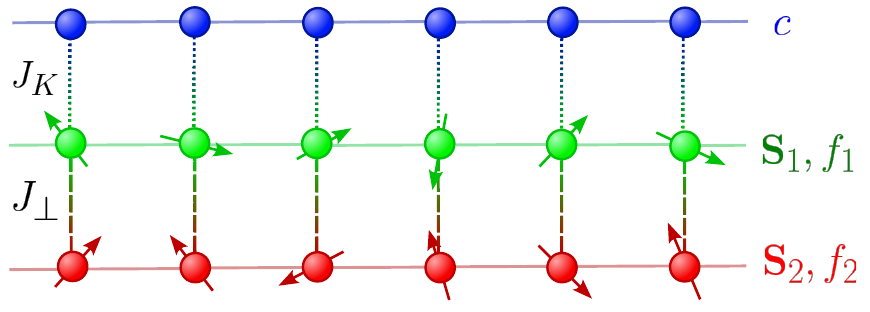}
       \caption{Ancilla model to describe a single band Hubbard model of electrons $c$ on a square lattice. The ancilla are $S=1/2$ spins ${\bm S}_1$, ${\bm S}_2$ in bilayer square lattice antiferromagnet. The Kondo coupling between the electrons and first layer of spins is denoted by a dotted line with coupling $J_K$, the antiferromagnetic coupling between the first and second layer of spins is denoted with a dashed line and $J_\perp$.}
    \label{fig:2}
\end{figure}

Fermi surfaces which do not enclose the Luttinger volume are relatively easy to obtain in two-band  Kondo lattice models \cite{FLS1,FLS2,Arun04} coupling a single band of conduction electrons to a second band of localized spins. One assumes the `small' Luttinger-volume-violating Fermi surface is formed by the conduction electrons alone, while the localized spins form a fractionalized spin liquid. However, it is much more difficult to obtain a systematic model of a Luttinger-volume-violating Fermi surface in a single-band Hubbard-like model. Here we employ the ancilla approach introduced in Ref.~\cite{Ancilla1} to describe a single-band Hubbard model of electrons $c$ with on-site repulsion $U$, which has successfully described photoemission \cite{Mascot22} and polaronic correlations \cite{Koepsell21,Yasir24,Shackleton24} in the pseudogap metal. After a Schrieffer-Wolff-type transformation \cite{Mascot22}, the Hubbard interaction is realized by coupling otherwise free $c$ electrons to an insulating bilayer square lattice antiferromagnet of spin-1/2 moments ${\bm S}_1$ and ${\bm S}_2$ (see Fig.~\ref{fig:2}).
For large rung-exchange $J_\perp$ in the bilayer antiferromagnet, the ancilla spins can be eliminated by a canonical transformation, yielding a Hubbard interaction $U \sim J_K^2/J_\perp$ between the $c$ electrons, where $J_K$ is the Kondo interaction between the $c$ electrons and the ${\bm S}_1$ spins. 
When $J_\perp \gg J_K$, the ${\bm S}_{1,2}$ spins form a trivial rung-singlet state, while the $c$ electrons form a Luttinger-volume `large' Fermi surface. The non-Luttinger volume, FL* pseudogap metal hole pocket phase of interest in this paper is obtained at smaller $J_\perp$, when the $c$ electrons are Kondo-screened by the ${\bm S}_1$ spins, while the ${\bm S}_2$ spins form a $\pi$-flux spin liquid.

The ${\bm S}_1$ and ${\bm S}_2$  spins can be represented by spin-1/2 fermionic partons $f_{1,2}$ via 
\begin{align} 
{\bm S}_1=f_1^\dagger\boldsymbol{\sigma}f_1 \quad , \quad {\bm S}_2=f_2^\dagger\boldsymbol{\sigma}f_2\,, 
\end{align}
with $\boldsymbol{\sigma}$ the Pauli matrices. In the ancilla mean-field theory \cite{Ancilla1}, the pseudogap regime is characterized by a finite hybridization between the $c$-electrons and the $f_1$-fermions, and a decoupled $\pi$-flux spin liquid for the $f_2$ spinons, with an emergent SU(2) gauge field. The (condensed) hybridization field $\phi$ can be introduced decoupling the Kondo interaction between the $c$-electrons and the $f_1$-fermions. The term $J_\perp{\bm S}_1\cdot{\bm S}_2$ can be further decoupled introducing a \textit{chargon} field $B=(B_1,B_2)$. In the pseudogap regime the chargon field carries a unit electromagnetic charge and, along with the $f_2$ spinons, it lives in the fundamental representation of the emergent SU(2) gauge group. The condensation of $B$ will fully confine (Higgs) the SU(2) gauge field and, depending on the exact form of the condensate, it leads to different symmetry broken states, such as $d$-wave superconductivity, CDW, or current loop orders~\cite{PNAS_pseudo}. There are no unbroken gauge symmetries once $B$ is condensed, and so a mean-field treatment is reasonable, as in the heavy Fermi liquid state of Kondo lattice models.

\subsection*{Mean-field Hamiltonian}

In this paper, we focus on a mean-field ansatz for $B$ that induces \textit{bi-directional} period-4 CDW order. Additional results for other periods CDW are presented in Sec.~\ref{app:period6} of the Supplementary Information (SI). The mean-field Hamiltonian reads
\begin{equation}\label{eq: MF Hamiltonian}
    \begin{split}
        \mathcal{H} = &
        \sum_{{\bm i},{\bm j}}\left[t^c_{{\bm i}{\bm j}}c^\dagger_{{\bm i}}c_{{\bm j}}+t^f_{{\bm i}{\bm j}}f^\dagger_{1,{\bm i}}f_{1,{\bm j}}\right]\\
        +&\sum_{{\bm i}}\left[\phi\, c^\dagger_{{\bm i}}f_{1,{\bm i}}+\mathrm{H.c.}\right]
         +i J\sum_{\langle {\bm i},{\bm j}\rangle}f^\dagger_{2,{\bm i}}e_{{\bm i}{\bm j}}f_{2,{\bm j}}\\
         +&i\sum_{{\bm i}}\left[B_{1{\bm i}}\,f^\dagger_{2,{\bm i}}f_{1,{\bm i}}-B_{2{\bm i}}f_{2,{\bm i}}(i\sigma^2)f_{1,{\bm i}}+\mathrm{H.c.}\right]\\ 
         +&ig\sum_{{\bm i}}\left[B_{1{\bm i}}\,f^\dagger_{2,{\bm i}}c_{{\bm i}}-B_{2{\bm i}}f_{2,{\bm i}}(i\sigma^2)c_{{\bm i}}+\mathrm{H.c.}\right]\,, 
    \end{split}
\end{equation}
where ${\bm i}=(x,y)$, $t^c_{{\bm i}{\bm j}}$ and $t^f_{{\bm i}{\bm j}}$ are hopping parameters that can be obtained fitting angular resolved photoemission (ARPES) data~\cite{Mascot22}, and $e_{{\bm i}{\bm j}}$ are the $\pi$-flux hoppings, defined in \appref{app: MF ansatz} and in Ref.~\cite{PNAS_pseudo}. Fluctuations of the hybridizaton field $\phi$ can induce a direct coupling $g$ of the chargons and $f_2$ spinons to the $c$-electrons. For this reason we have included the last line of \eqref{eq: MF Hamiltonian} in our Hamiltonian. We make an ansatz for $B$ such that the induced CDW takes the form
\begin{equation}\label{eq: chargon CDW}
    B^\dagger_{\bm i}B_{\bm i} = |b|^2 \left(1-\frac{(-1)^x+(-1)^y}{2}\right)\cos^2(qx)\cos^2(qy)\,,
\end{equation}
where $b$ is a parameter quantifying the strength of CDW order, and $q=\pi/4$, and no additional superconducting or time-reversal-symmetry-breaking orders are induced (see \appref{app: MF ansatz}). The Fourier transform of \eqref{eq: chargon CDW} has delta functions at a number of wavevectors (including $(\pi, 0)$ and $(0,\pi)$ listed below \eqref{seq: CDW form} of the SI), apart from the primary CDW wavevectors $(\pm 2 q , 0)$, $(0, \pm 2q)$. These appear as a consequence of the confinement transition associated with the condensation of $B$ that leads to the CDW phase, and are, in principle, an observable consequence of our theory. However, we expect that these additional wavevectors will have distinct renormalizations from SU(2) gauge fluctuations (not included in our mean-field theory here),  which are expected to suppress the additional wavevectors relative to the primary CDW wavevectors; this is an important questions for further research which accounts for the gauge fluctuations more completely. In \appref{app: CDW profiles}, we show the charge and bond density profiles induced by the mean-field ansatz~\eqref{eq: chargon CDW}.

Another distinct possible CDW phase of the ancilla model is a so-called CDW* phase, in which the $c$-electron density is spatially dependent but the SU(2) gauge field remains deconfined. This can be achieved setting $B=0$ in Hamiltonian~\eqref{eq: MF Hamiltonian} and adding a modulated potential $V\sum_{\bm{i}}[\cos(Qx)+\cos(Qy)]c^\dagger_{{\bm i}}c_{{\bm i}}$, with $Q=\pi/2$, in the top layer. The ancilla description of the CDW order of such a phase is similar to the one considered in Ref.~\cite{Zhang_2016}, with the difference that we added a set of self-consistently calculated Lagrange multipliers coupled to the $f_1$ fermions to keep their density uniform and equal to 1 (see \appref{SMsec: MF Hamiltonian}); see also Ref.~\cite{Chowdhury14}. (We adopted this procedure also for the $f_2$ fermions in the CDW phase.)  However, these earlier approaches do not explicitly account for the spinon excitations in the CDW* phase, which are required to be present by general arguments \cite{FLS2}, and which are automatically included in the ancilla approach.

\section{Fermi surfaces}
\label{sec:fermi}

\begin{figure*}[t]
    \includegraphics[width=\textwidth]{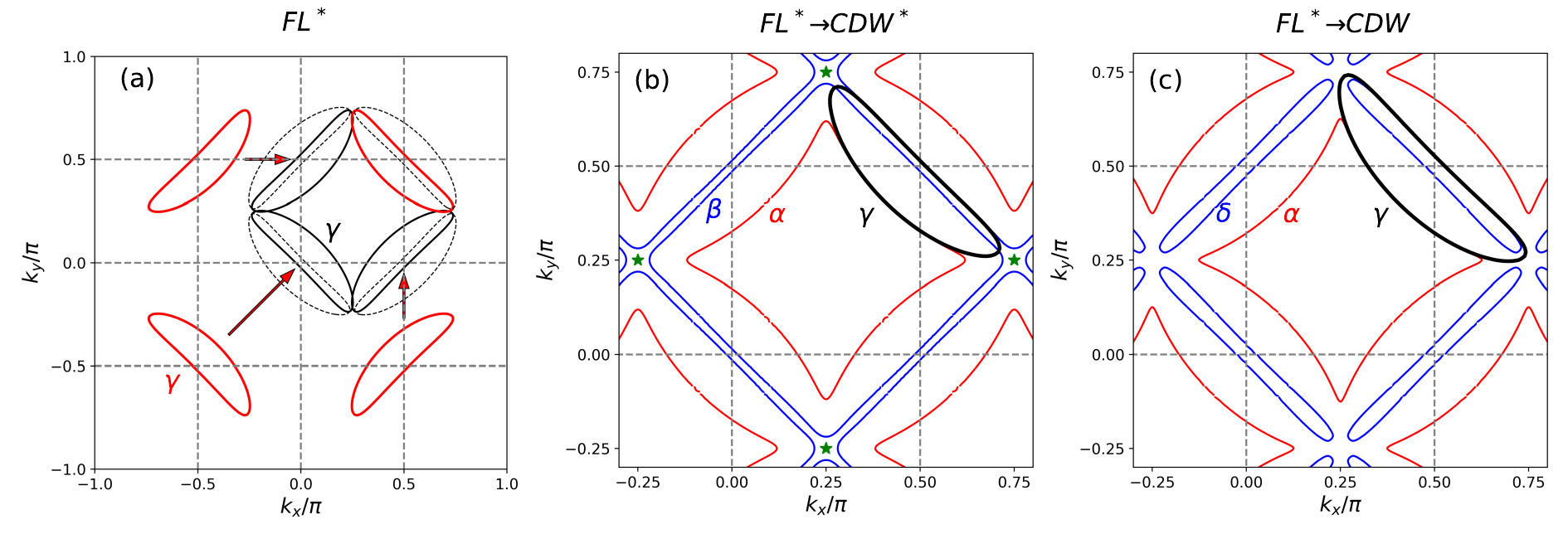}
       \caption{Computed Fermi surface reconstruction in the (a) FL* pseudogap metal, (b) FL*$\rightarrow$CDW*,  and (c) FL*$\rightarrow$CDW  phases of the ancilla model; (c) is our proposal for cuprate quantum oscillations. In (a) the four $\gamma$ hole pockets (enclosing non-Luttinger area $p$) are brought close to each other by integer combinations of the CDW momenta $(\frac{\pi}{2},0)$ and $(0,\frac{\pi}{2})$. The dashed lines represent the area obtained by shifting the hole pockets by higher harmonics. (b) In the CDW* phase one has $\alpha$ and $\beta$ Fermi pockets, as in Ref.~\cite{Zhang_2016}. The black $\gamma$ pockets is the same as the red $\gamma$ pocket in (a). A co-existing $\pi$-flux spin liquid is present in CDW*, and its spinons have nodal points located at the green stars, which are shifted by multiples of {\it half} the CDW wavevectors (see \appref{app: MF ansatz}). (c) In the CDW phase, the $\pi$-flux spinons hybridize with the $f_1$-fermions and the $c$-electrons, thereby disrupting the $\beta$ pocket and leaving four $\delta$ pockets of small volume. The $\alpha$ electron pocket remains instead unaffected. The dashed gray lines represent the boundaries between different reduced Brillouin zones.}
    \label{fig:3}
\end{figure*}
\begin{figure}[t]
    \includegraphics[width=3.2in]{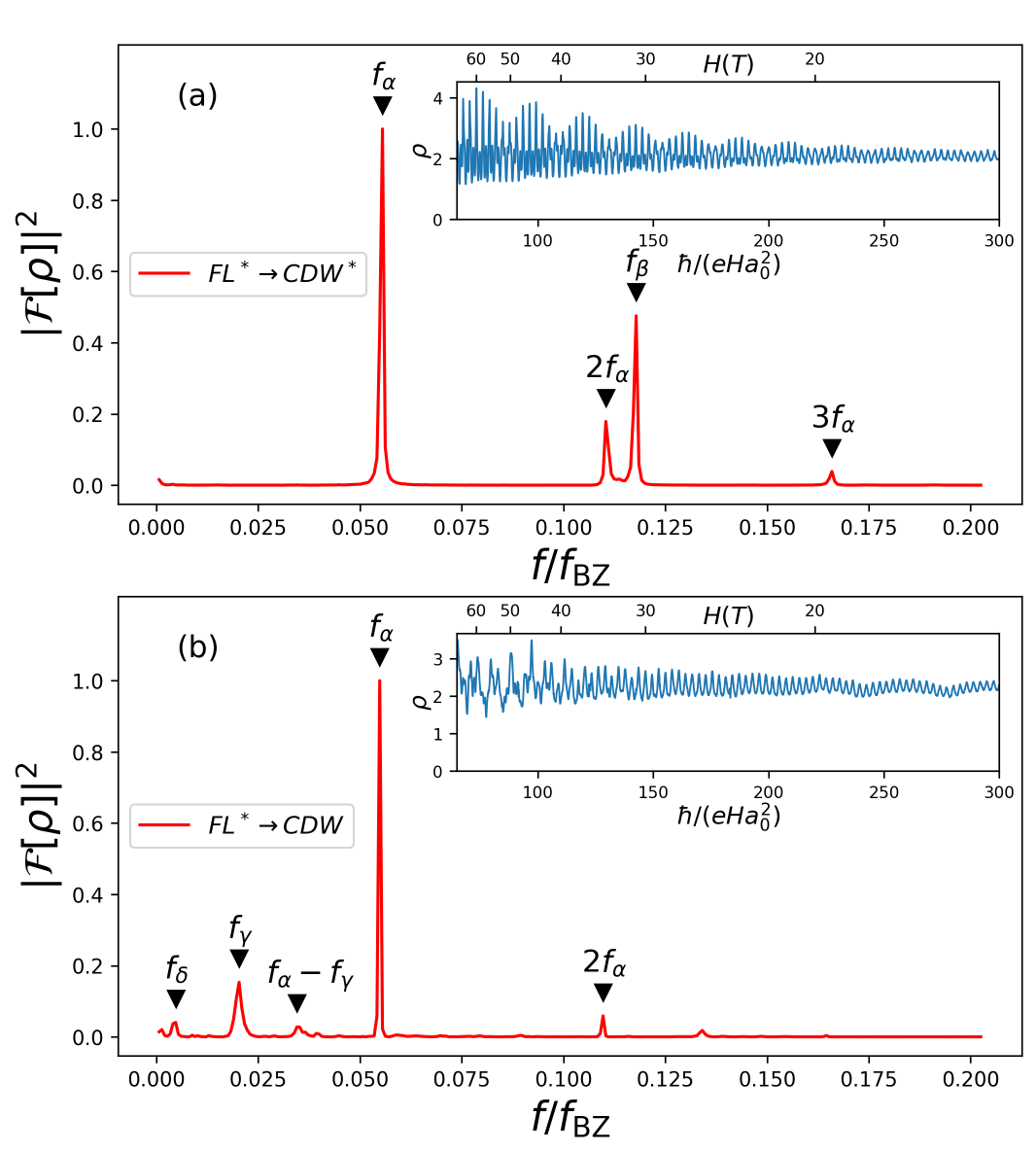}
       \caption{Quantum oscillations in the (a) FL*$\rightarrow$CDW* and (b) FL*$\rightarrow$CDW phases of the ancilla model; (b) is our result for the cuprates. The insets show oscillations of the density of states $\rho$ as a function of the inverse magnetic field, whereas the main panels display their Fourier spectrum. The labeling of the peaks corresponds to the Fermi surface sheets indicated in Fig.~\ref{fig:3} in panels (b) and (c), respectively. Both $\rho$ and its Fourier spectrum $|\mathcal{F}[\rho]|^2$ are measured in arbitrary units.}
    \label{fig:4}
\end{figure}

The Fermi surfaces in CDW phase descending from the large Fermi surface Fermi liquid (FL) have been computed earlier \cite{Allais14} (for FL$\rightarrow$CDW only $t^c$ and $V$ are non-zero). Here we obtain the distinct Fermi surfaces for FL*$\rightarrow$CDW and FL*$\rightarrow$CDW*, which describe reconstruction of the hole pockets in Fig.~\ref{fig:1} by CDW order. In \eqref{eq: MF Hamiltonian}, we choose $t^c_{{\bm i}{\bm j}}$ and $t^f_{{\bm i}{\bm j}}$ as in Refs.~\cite{Mascot22,Christos23}, $\phi=0.1$ eV, $J=0.14$ eV, and fix the $c$-electron density to $n_c=0.8425$, which, in absence of CDW order, gives $Q_\mathrm{cdw}\sim \pi/2$ (see Fig.~\ref{fig:1}). We calculate the CDW bandstructure in the \textit{reduced} Brillouin zone (BZ) ${(k_x,k_y)\in[0,\pi/2)}$, but, for visualization purposes, we \textit{unfold} the resulting Fermi surfaces to an enlarged BZ ${(k_x,k_y)\in[-\pi/4,3\pi/4)}$. The Fermi surfaces in the reduced BZ are shown in \appref{app:FS}.

In Fig.~\ref{fig:3}, we show the Fermi surface reconstruction in the CDW* (panel b) and CDW (panel c) states. In panel (a) of Fig.~\ref{fig:3} we show how the four hole pockets appear when the $\gamma$ pockets are shifted by wavevectors of the form $(\frac{\pi}{2}n_x,\frac{\pi}{2}n_y)$, $n_x,n_y\in \mathbb{Z}$. In panel (b) of Fig.~\ref{fig:3}, we show the Fermi surfaces in the CDW* state for $V\simeq 0.02$ eV. The CDW potential opens a gap at the points where the shifted pockets of Fig.~\ref{fig:3}a cross, leaving two electron pockets, $\alpha$ and $\beta$, both centered around $(\pi/4,\pi/4)$, and whose spectral weight comes mostly from the $c$- and $f_1$-fermions, respectively. The $\pi$-flux spin liquid remains decoupled and displays nodal points at the locations marked with green stars. In Fig.~\ref{fig:3}c, we show the Fermi surfaces in a CDW state, obtained for $|b|\simeq0.1$ eV, $g=1$, and with an additional modulated potential for the $c$-electrons with $V\simeq 0.01$ eV. We observe that the presence of a finite $b$ inflates the $\pi$-flux nodes to proper Fermi surfaces, which then hybridize with the backside of the hole pockets, leaving two different kind of Fermi surfaces: an $\alpha$ pocket, similar to the CDW* phase, and two elongated hole pockets, which we denote by $\delta$, and which enclose a volume corresponding to a tiny fraction of the full Brillouin zone. Note that we have added a small modulated potential in Fig.~\ref{fig:3} only to get rid of very small hole pockets (much smaller than the $\delta$ pockets) appearing in the vicinity of the nodes of the quantum spin liquid dispersion. Also note that a $g\sim 1$ is essential to obtain the $\delta$-pockets instead of a larger $\beta$-pocket. In fact, for $g=0$ and small $b$, one gets similar Fermi surfaces as in Fig.~\ref{fig:3}c, but further increasing $b$ closes the gap between the $\delta$-pockets again, giving back a $\beta$-pocket.

\section{Quantum Oscillations}
\label{sec:qo}

We now turn our attention onto how the Fermi surfaces displayed in Fig.~\ref{fig:3}(b)-(c) for FL*$\rightarrow$CDW* and FL*$\rightarrow$CDW show up in quantum oscillations in a magnetic field \cite{SuchitraCyril,Proust_review} (quantum oscillations for FL$\rightarrow$CDW \cite{Allais14} do not agree with observations). We couple Hamiltonian~\eqref{eq: MF Hamiltonian} to a uniform magnetic field by multiplying the hopping terms $t^c_{{\bm i}{\bm j}}$ and $t^f_{{\bm i}{\bm j}}$ by a Peierls phase
\begin{equation}\label{eq: Peierls phase}
    e^{i{\bm A_{({\bm i}+{\bm j})/2}}\cdot({\bm i}-{\bm j})}\,.    
\end{equation}
In the pseudogap phase, the condensation of the hybridization field $\phi$ gives the $f_1$ fermions an electromagnetic charge, which is why also the hoppings $t^f_{{\bm i}{\bm j}}$ must acquire a Peierls phase. In the CDW regime, when $B$ condenses, the $f_2$ fermions become charged under the electromagnetic U(1) symmetry. For this reason, in the calculations in the CDW phase (but not in the CDW* phase) we also multiply $e_{{\bm i}{\bm j}}$ by~\eqref{eq: Peierls phase}. We choose a gauge in which ${\bm A}_{\bm i}= Hx (0,1,0)$, with $H$ the magnetic field measured in units of $H_0=\hbar/(e a_0^2)$, with $e$ the elementary charge, $\hbar$ the reduced Planck constant, and $a_0\simeq 3.86\r{A}$ the lattice spacing. As a function of $H$, we have computed the density of states at the Fermi level of Hamiltonian~\eqref{eq: MF Hamiltonian} on a system with $N_x=2000$ sites and open boundaries in the $x$-direction, and $N_y=96$ sites and periodic boundaries in the $y$-direction (see \appref{app:QO} for details on the numerical procedures).

In Fig.~\ref{fig:4}, we show the quantum oscillations of the density of states $\rho$ as a function of $1/H$ (insets), as well as their Fourier spectra (main plots), both in the CDW* (panel a) and CDW (panel b) phases. The parameters chosen for Fig.~\ref{fig:4} are exactly the same as in Fig.~\ref{fig:3}. Plots for different choices of the parameters are shown in \appref{app:additional}. 

In the CDW* phase (Fig.~\ref{fig:4}a), we observe a main peak in the Fourier transform of $\rho(1/H)$ occurring at frequency $f_\alpha\simeq 0.055 f_\mathrm{BZ}\sim1500$ T ($f_\mathrm{BZ}=(2\pi)^2 H_0/(2\pi)\simeq 27.6$ kT), corresponding to the $\alpha$ pocket. However, additional peaks appear, with heights that cover a significant fraction of the height of the main peak. Aside from higher harmonics of the main frequency $f_\alpha$, we observe a sizeable peak ($\simeq$ 47.5\% of the height of the peak at $f_\alpha$) at the frequency $f_\beta\simeq 0.12 f_\mathrm{BZ}\sim3300$ T, corresponding to the $\beta$ pocket of Fig.~\ref{fig:3}b. 

Differently, in the CDW phase (Fig.~\ref{fig:4}b), the peak at $f_\alpha$ is much more pronounced. We observe a tiny peak ($\simeq$ 4\% of the $\alpha$ peak) at $f_\delta=0.004 f_\mathrm{BZ}$, corresponding to the $\delta$ pocket in Fig.~\ref{fig:3}c. A magnetic breakdown frequency emerges at $f_\gamma/f_\mathrm{BZ}\simeq 0.02$, corresponding to $(1-n)/8$, that is the volume of the hole pockets in absence of CDW order (denoted by $\gamma$ in Fig.~\ref{fig:3}). Other lower peaks appear at $f=f_\alpha-f_\gamma$ and $f=2f_\alpha$. 
\section{Results for period 6 CDW and CDW*}
\label{sec: period 6}

In this section we present results obtained for period 6 CDW* and CDW phases. Within our formalism, we can only obtain density modulations with \textit{even} periods, as the wavevectors $(\pi,0)$ and $(0,\pi)$ are always present in the CDW order parameter (see Ref.~\cite{PNAS_pseudo} and \appref{app: MF ansatz} for details). It is possible, however, that fluctuations beyond mean-field will remove the CDW components at $(\pi,0)$ and $(0,\pi)$. In \appref{app: CDW profiles}, we show the charge and bond density profiles for a period 6 CDW phase.

To study a period 6 CDW and CDW* phase, we start from Hamiltonian~\eqref{eq: MF Hamiltonian} and make an ansatz for $B$ such that \eqref{eq: chargon CDW} still holds but with $q=\pi/6$. We fix the $c$-electron density to $n=0.945$, so that the nesting vector $Q_\mathrm{cdw}$ in Fig.~\ref{fig:1} equals $(2\pi)/3$. Moreover, we add a modulated potential in the top layer, $V\sum_{\bm{i}}[\cos(Qx)+\cos(Qy)]c^\dagger_{{\bm i}}c_{{\bm i}}$, with $Q=(2\pi)/3$. Additional details are provided in \appref{app:period6}, where we also show the Fermi surfaces in the reduced Brillouin zone for a CDW* and a CDW phase. Also in this case, we find an $\alpha$ and a $\beta$ pocket in the CDW* state, while in the CDW phase the hybridization of the $\beta$ pocket with the $f_2$ spinons disrupts it into two elongated $\delta$ pockets (see Fig.~\ref{figSM:5} in \appref{app:period6}). 

When calculating the quantum oscillations of the density of states in the period 6 CDW* phase, we get peaks at the frequencies corresponding to the volumes of the $\alpha$ and $\beta$ pockets, similar to Fig.~\ref{fig:4}a. Differently, in the CDW phase the $\beta$ pocket peak is strongly suppressed, but, unlike the period 4 case, it is still present with a weak intensity, possibly due to magnetic breakdown processes. The $\alpha$ frequency obtained in this case is about 650 T, which is in much better agreement with the experimentally observed frequencies (500 T to 900 T, depending on the material and doping level). These results are displayed in Fig.~\ref{figSM:6} in \appref{app:period6}.

\section{Discussion}
\label{sec:conc}

We summarize the manner in which our computation reconciles experimental observations, as claimed in the introduction.\\
({\it i\/}) Our FL* $\gamma$ Fermi surfaces in Fig.~\ref{fig:3}a are very similar to those in Ref.~\cite{Mascot22}, which matched zero field photoemission data in both the nodal and anti-nodal regions of the Brillouin zone.\\
({\it ii\/}) We have shown how backside problem (B) is resolved by the FL*$\rightarrow$CDW transition, as the spinons of the FL* pseudogap metal remove the $\beta$ Fermi surface in Figs.~\ref{fig:3}b and \ref{fig:4}a, while preserving the Harrison-Sebastian electron pocket $\alpha$ in Figs.~\ref{fig:3}c and \ref{fig:4}b. The new $\delta$ pockets in the CDW phase in Fig.~\ref{fig:3}c show up only at very low frequencies in the quantum oscillations in Fig.~\ref{fig:4}b, making them very difficult to detect over the available field range. Moreover, the $\delta$ pockets are small and not associated with a significant electron density, and so the chemical potential oscillations in a magnetic field will be those of the $\alpha$ pocket alone, as is observed \cite{Sebastian11,Greven16}.\\
({\it iii\/}) Although they are not manifest in the quantum oscillations, the two small $\delta$ pockets will have significant consequences for the specific heat at high field. Indeed, the combination of the single $\alpha$ pocket and the two $\delta$ pockets yields just the required factor of 3 enhancement  in the linear-in-temperature co-efficient of the specific heat observed by Girod {\it et al.\/} in HgBa$_{2}$CuO$_{4+\delta}$ \cite{Girod20} (assuming the effective masses of the $\alpha$ and $\delta$ pockets are the same). There is an enhancement of around a factor of 2 in
YBa$_2$Cu$_3$O$_{6.56}$ \cite{Girod20,Boebinger11}, and this is possibly connected to the presence of bilayers.

We have not addressed issues related to bi- versus uni-directional CDW order \cite{Kivelson19}, but suggest that the resolution may lie in charge order that has anisotropic strengths \cite{Allais14}.

The spinons and chargons employed in our theory of quantum oscillations should also help resolve backside problem (A)---the presumed completion of the Fermi arcs into $\gamma$ hole pockets in the higher temperature pseudogap metal, supported by angle-dependent magnetoresistance (ADMR) observations \cite{ADMR}. Quantum and thermal fluctuations of low energy chargons should reduce the backside spectral intensity, just as condensation of chargons has resolved backside problems in the (B) quantum oscillations and (C) $d$-wave superconductor. Such effects could also resolve the issues with interpreting ADMR raised in Ref.~\cite{Musser22} 

\subsection{Direct detection of spinons}

Given the central role of the spinons in resolving backside problems (B) and (C), it would be of great interest to detect the spinons in neutron scattering. There has been recent progress in neutron scattering detection of spinons in triangular lattice Mott insulators \cite{Batista24,Haravifard24}. 
For the insulating, undoped square lattice, there is long-range N\'eel order, but signatures of spinons have been claimed in the higher energy neutron scattering \cite{Hayden10,Ronnow15}.
Detection on the doped square lattice without magnetic order will require accurate theoretical computations of the higher energy dynamic spin structure factor from spinons at the N\'eel-valence bond solid transition \cite{NRSS89,DQCP3,Zhou:2023qfi,Meng23,Chester:2023njo,Sandvik24,YZY18}. 
At lower energies, the influence of the doping is likely to be more appreciable, and approaches using confined degrees of freedom \cite{Vojta06,AlexN23} are more appropriate.

The mean-field theory of fermionic spinons has gapless Dirac spinons at 
wavevector $(\pi/2, \pi/2)$, and this predicts spinon continua of equal intensity at $(\pi,\pi) $ and $(\pi, 0)$. However, it is now possible to go well beyond the free spinon theory \cite{Iqbal19}, and include the consequences of the SU(2) gauge fluctuations. We can accurately estimate the magnitude of neutron scattering from spinons at various wavevectors in the insulating quantum-critical spin liquid from the remarkable fuzzy sphere results of Zhou {\it et al.} \cite{Zhou:2023qfi}.
If the scaling dimension of the spin operator at a wavevector ${\bf q}$ is $\Delta_{\bf q}$, then the Fourier transform implies that the zero temperature dynamic structure factor, $S({\bf q}, \omega)$, at ${\bf q}$ diverges as 
\begin{align}
    S({\bf q}, \omega) \sim \omega^{2 \Delta_{\bf q} -3}\,.
\end{align} 
We obtain estimates of $\Delta_{\bf q}$ in Table II of Zhou {\it et al.} \cite{Zhou:2023qfi}. 
At ${\bf q} = (\pi, \pi)$, we use the SO(5)-fundamental operator $\phi$ as 3 components of it correspond to N\'eel order. We therefore obtain 
\begin{align}
    \Delta_{(\pi, \pi)} = 0.585\,,
\end{align}
yielding a strong dynamic spectrum scaling as $\omega^{-1.83}$. 

At ${\bf q} = (\pi, 0)$, a first attempt is to employ the operator with spin $S=1$ and momentum $(\pi,0)$ which is obtained from the fusion of the N\'eel order (which has $S=1$ and momentum $(\pi, \pi)$) and valence bond solid (VBS) order (which has $S=0$ and momentum $(0, \pi)$). The VBS order is also a component of $\phi$, and so we have to consider the fusion of $\phi$ with itself. This yields the operator $T$ of Zhou {\it et al.} \cite{Zhou:2023qfi}: from their Table II, its scaling dimension yields
$\Delta_{(\pi, 0)} = 1.458$, and
a dynamic spectrum scaling as $\omega^{-0.084}$, which is a very weak divergence. However, a more careful analysis shows that the scaling at $(\pi, 0)$ is even weaker. The VBS order is odd under certain reflections about lattice sites, and we need an operator which is even under all such reflections. So we have to consider the fusion of $\phi$ with $\partial_{x,y} \phi$ \cite{YZY18}. This yields the first descendant of $T$, with dimension $2.458$, and also the SO(5) current $J_\mu$ with dimension $2$. Choosing the smaller dimension, we conclude
\begin{align}
    \Delta_{(\pi, 0)} = 2,
\end{align}
yielding a dynamic spectrum scaling as $\omega$. This extremely weak scaling explains the absence of any signal in neutron scattering at $(\pi, 0)$ \cite{Dai01,Keimer00,Keimer07}. 

For completeness, we note that the total magnetization also scales as $J_\mu$, and so $\Delta_{(0,0)} = 2$.

These results imply that SU(2) gauge fluctuations strongly renormalize the free Dirac spinon predictions for $S({\bf q}, \omega)$. The dominance of spin fluctuations at $(\pi,\pi)$ indicates that the $\mathbb{CP}^1$ formulation of the same spin liquid may be a better starting point for computations of $S({\bf q}, \omega)$ \cite{AlexN23}. Nevertheless, the Dirac fermion formulation has been useful for addressing the nature of the fermionic excitations for quantum oscillations in the present paper, and of the Bogoliubov quasiparticles in the superconducting state \cite{Christos23}. 

On the experimental side, the remarkable observations of Refs.~\cite{Keimer11,Keimer22,Hayden19} indeed show high energy spin fluctuations in the pseudogap regime of the doped system without antiferromagnetic order. These spin fluctuations are near $(\pi, \pi)$, as expected from the theoretical considerations above.
They appear as remnants of the spin waves of the antiferromagnet, and have been interpreted as damped paramagnons. However, there is no large Fermi surface at this doping, and so it does not seem reasonable that the large spectral weight of ``intense paramagnon excitations'' \cite{Keimer11} can be due to particle-hole fluctuations on the observed small Fermi surface. Moreover, a spin-wave interpretation is only valid at low energies in the antiferromagnetically ordered state. We argue that the  natural interpretation of the signal is that it is a spinon continuum, similar to that computed in Ref.~\cite{YZY18}. This aligns the observations of Refs.~\cite{Keimer11,Keimer22,Hayden19} with the spinons required by the small non-Luttinger-volume Fermi surface of the FL* description of the pseudogap metal. More precise theoretical computations and experimental observations should help settle the issue.

\subsection*{Acknowledgements}

We thank Bill Atkinson, Greg Boebinger, Yin-Chen He, Bernhard Keimer, Steve Kivelson, Matthieu Le Tacon, Zhu-Xi Luo, Zi Yang Meng, Cyril Proust, Brad Ramshaw, Anders Sandvik, Mathias Scheurer, Henry Shackleton, Louis Taillefer, Jia-Xin Zhang, and Ya-Hui Zhang for valuable discussions. This research was supported by the U.S. National Science Foundation grant No. DMR-2245246, by the Gordon and Betty Moore Foundation’s EPiQS Initiative Grant GBMF8683, and by the Simons Collaboration on Ultra-Quantum Matter which is a grant from the Simons Foundation (651440, S.S.). P.M.B. acknowledges support by the German National Academy of Sciences Leopoldina through Grant No. LPDS 2023-06.

\manuallabel{app:period6}{{6}}
\manuallabel{figSM:6}{{S6}}
\manuallabel{figSM:5}{{S5}}
\manuallabel{app: MF ansatz}{{1}}
\manuallabel{seq: CDW form}{{S2}}
\manuallabel{app: CDW profiles}{{7}}
\manuallabel{SMsec: MF Hamiltonian}{{2}}
\manuallabel{app:FS}{{4}}
\manuallabel{app:QO}{{3}}
\manuallabel{app:additional}{{5}}

\bibliography{refs}

\newpage


\section*{Supplementary material}

with $m$ and $N$ mutually prime positive integers, satisfying $m < N/2$. This choice limits us to the study of *commensurate* CDW order. The *incommensurate* limit can be studied by approximating $q$ as in Eq. (A5), with $m$ an $N$ suitably chosen to satisfy the desired degree of accuracy. In the following, we will assume $N$ to be even. This choice ensures that all Fourier modes of the CDW in Eq. (A2) can be expressed as $n_x(2q,0) + n_y(0,2q)$ modulo a reciprocal lattice vector, with $n_x, n_y \in \{0, 1, \ldots N-1\}$.

For future convenience, we re-write Ansatz (A1) in the form

$$B_{1,\boldsymbol{i}} = e^{i\left(\frac{\pi}{2}-q\right)(x+y)} \sum_{n_x,n_y=0}^{N-1} b_{n_x,n_y} e^{2iq(n_x\cdot x + n_y\cdot y)}. \tag{A6}$$

$b_{n_x,n_y}$ can be obtained comparing (A6) with (A1a).

### Appendix B: Mean-field theory of bi-directional CDW

Inserting expression (A1) into Eq. (2) of the main text, and Fourier transforming, we obtain the following mean-field Hamiltonian

$$\begin{aligned}
\mathcal{H} = & \sum_{\mathbf{k},\sigma} \left[ \epsilon_{\mathbf{k}}^c c_{\mathbf{k},\sigma}^\dagger c_{\mathbf{k},\sigma} + \epsilon_{\mathbf{k}}^f f_{1,\mathbf{k},\sigma}^\dagger f_{1,\mathbf{k},\sigma} + 2J\cos(k_x - q) f_{2,\mathbf{k},\sigma}^\dagger f_{2,\mathbf{k},\sigma} + 2J\cos(k_y - q) f_{2,\mathbf{k}+\mathbf{Q},\sigma}^\dagger f_{2,\mathbf{k},\sigma} \right] \\
& + \sum_{\mathbf{k},\sigma} \left[ \phi\, c_{\mathbf{k},\sigma}^\dagger f_{1,\mathbf{k},\sigma} + \text{H.c.} \right] + i \sum_{\mathbf{k},\sigma} \sum_{n_x,n_y=0}^{N-1} \left[ b_{n_x,n_y} f_{1,\mathbf{k},\sigma}^\dagger f_{2,\mathbf{k}+2q(n_x,n_y),\sigma} + \text{H.c.} \right] \\
& + ig \sum_{\mathbf{k},\sigma} \sum_{n_x,n_y=0}^{N-1} \left[ b_{n_x,n_y} c_{\mathbf{k},\sigma}^\dagger f_{2,\mathbf{k}+2q(n_x,n_y),\sigma} + \text{H.c.} \right],
\end{aligned} \tag{B1}$$

where $\sigma = \uparrow, \downarrow$ is the spin projection, $\epsilon_{\mathbf{k}}^c$ and $\epsilon_{\mathbf{k}}^f$ are the Fourier transforms of the real space hoppings $t_{ij}^c$ and $t_{ij}^f$, respectively, $\mathbf{Q} = (\pi, 0) \sim \frac{N}{2}(2q, 0)$, and we have redefined the $\pi$-flux spinons as $f_{2,\mathbf{k},\sigma} \to f_{2,\mathbf{k}-(\pi/2-q,\pi/2-q),\sigma}$. The explicit form of $\epsilon_{\mathbf{k}}^{c/f}$ is

$$\begin{aligned}
\epsilon_{\mathbf{k}}^{c/f} = & -2t_0^{c/f}(\cos k_x + \cos k_y) - 4t_1^{c/f} \cos k_x \cos k_y - 2t_2^{c/f}(\cos(2k_x) + \cos(2k_y)) \\
& - 4t_3^{c/f}(\cos(2k_x + k_y) + \cos(k_x + 2k_y)),
\end{aligned} \tag{B2}$$

with, as in Refs. [19, 44], $t_0^c = 0.22$ eV, $t_1^c = -0.034$ eV, $t_2^c = 0.036$ eV, $t_3^c = -0.007$ eV, $t_0^f = -0.1$ eV, $t_1^f = 0.03$ eV, $t_2^f = 0.01$ eV, $t_3^f = 0$ eV. We chose the hybridization field $\phi$ to equal 0.1 eV. These parameters have been obtained by fitting photoemission data for the compound Bi2201 ($\text{Pb}_{0.55}\text{Bi}_{1.5}\text{Sr}_{1.6}\text{La}_{0.4}\text{CuO}_{6+\delta}$) [19].

The $f_1$- and $f_2$-spinons, living in the middle and bottom layer, respectively, must satisfy the constraints

$$\sum_\sigma f_{1,\boldsymbol{i},\sigma}^\dagger f_{1,\boldsymbol{i},\sigma} = 1 \quad \forall \boldsymbol{i}, \tag{B3a}$$

$$\sum_\sigma f_{2,\boldsymbol{i},\sigma}^\dagger f_{2,\boldsymbol{i},\sigma} = 1 \quad \forall \boldsymbol{i}. \tag{B3b}$$

We enforce these constraints on average by adding two sets of Lagrange multipliers to the Hamiltonian, as well as a chemical potential for the $c$-electrons to keep their density fixed:

$$\mathcal{H} \to \mathcal{H} - \sum_{\boldsymbol{i},\sigma} \left[ \mu^c c_{\boldsymbol{i},\sigma}^\dagger c_{\boldsymbol{i},\sigma} + \sum_{n_x,n_y=0}^{N-1} \left( \mu_{n_x,n_y}^{f_1} f_{1,\boldsymbol{i},\sigma}^\dagger f_{1,\boldsymbol{i},\sigma} + \mu_{n_x,n_y}^{f_2} f_{2,\boldsymbol{i},\sigma}^\dagger f_{2,\boldsymbol{i},\sigma} \right) e^{2iq(n_x x + n_y y)} \right]. \tag{B4}$$

Finally, we add a modulated potential in the top layer,

$$\mathcal{H} \to \mathcal{H} + V \sum_{\boldsymbol{i},\sigma} [\cos(2qx) + \cos(2qy)] c_{\boldsymbol{i},\sigma}^\dagger c_{\boldsymbol{i},\sigma}. \tag{B5}$$

Defining the $3N^2$-component spinor

$$\Psi_{\mathbf{k},\sigma} = \begin{pmatrix} \widetilde{\Psi}_{\mathbf{k},\sigma}^{0,0} & \widetilde{\Psi}_{\mathbf{k},\sigma}^{0,1} & \dots & \widetilde{\Psi}_{\mathbf{k},\sigma}^{0,N-1} & \widetilde{\Psi}_{\mathbf{k},\sigma}^{1,0} & \dots & \widetilde{\Psi}_{\mathbf{k},\sigma}^{N-1,N-1} \end{pmatrix}^T, \tag{B6a}$$

$$\widetilde{\Psi}_{\mathbf{k},\sigma}^{n_x,n_y} = \begin{pmatrix} c_{\mathbf{k}+2q(n_x,n_y),\sigma} & f_{1,\mathbf{k}+2q(n_x,n_y),\sigma} & f_{2,\mathbf{k}+2q(n_x,n_y),\sigma} \end{pmatrix}, \tag{B6b}$$

we can cast Hamiltonian (B5) in the form

$$\mathcal{H} = \sum_{\mathbf{k},\sigma}{}' \sum_{\ell,\ell'=1}^{3N^2} \Psi_{\mathbf{k},\ell,\sigma}^\dagger \, \mathcal{H}_{\mathbf{k}}^{\ell\ell'} \, \Psi_{\mathbf{k},\ell',\sigma}, \tag{B7}$$

where the momentum sum $\sum_{\mathbf{k}}'$ is performed over the *reduced* Brillouin zone (BZ) $k_x, k_y \in [0, 2\pi/N)$.

For a fixed set of $b$, $g$, $V$ and for a fixed electron density, we self-consistently determine $\mu^c$, $\mu_{n_x,n_y}^{f_1}$ and $\mu_{n_x,n_y}^{f_2}$ imposing $n = \sum_{\mathbf{k},\sigma}\langle c_{\mathbf{k},\sigma}^\dagger c_{\mathbf{k},\sigma}\rangle$ and the constraints (B3) on average. Once convergence is achieved, we diagonalize $\mathcal{H}_{\mathbf{k}}^{\ell\ell'}$ to obtain the band structure in the CDW state.

Calculations in the CDW* state are performed setting $b=0$ and $\mu_{n_x,n_y}^{f_2} = 0$.

## Appendix C: Calculation of the Quantum oscillations

To compute the quantum oscillations of the density of states as a function of the inverse magnetic field $1/H$, we consider the Hamiltonian

$$\begin{aligned}\mathcal{H} =& \sum_{\boldsymbol{i},\boldsymbol{j}} \left[ t_{\boldsymbol{ij}}^c e^{i\boldsymbol{A}_{(\boldsymbol{i}+\boldsymbol{j})/2}\cdot(\boldsymbol{i}-\boldsymbol{j})} \, c_{\boldsymbol{i}}^\dagger c_{\boldsymbol{j}} + t_{\boldsymbol{ij}}^f e^{i\boldsymbol{A}_{(\boldsymbol{i}+\boldsymbol{j})/2}\cdot(\boldsymbol{i}-\boldsymbol{j})} \, f_{1,\boldsymbol{i}}^\dagger f_{1,\boldsymbol{j}} \right] \\ &+ \sum_{\boldsymbol{i}} \left[ \phi\, c_{\boldsymbol{i}}^\dagger f_{1,\boldsymbol{i}} + \text{H.c.} \right] + iJ \sum_{\langle\boldsymbol{i},\boldsymbol{j}\rangle} f_{2,\boldsymbol{i}}^\dagger \, e^{ie_{\text{qsl}}\boldsymbol{A}_{(\boldsymbol{i}+\boldsymbol{j})/2}\cdot(\boldsymbol{i}-\boldsymbol{j})} \, e_{\boldsymbol{ij}} \, f_{2,\boldsymbol{j}} \\ &+ i \sum_{\boldsymbol{i}} \sum_{n_x,n_y=0}^{N-1} \left[ e^{i(\frac{\pi}{2}-q)(x+y)} b_{n_x,n_y} \, e^{2iq(n_x\, x + n_y\, y)} \, f_{2,\boldsymbol{i}}^\dagger f_{1,\boldsymbol{i}} + \text{H.c.} \right] \\ &+ ig \sum_{\boldsymbol{i}} \sum_{n_x,n_y=0}^{N-1} \left[ e^{i(\frac{\pi}{2}-q)(x+y)} b_{n_x,n_y} \, e^{2iq(n_x\, x + n_y\, y)} \, f_{2,\boldsymbol{i}}^\dagger c_{\boldsymbol{i}} + \text{H.c.} \right] \\ &- \sum_{\boldsymbol{i},\sigma} \left[ \mu^c c_{\boldsymbol{i}}^\dagger c_{\boldsymbol{i}} + \sum_{n_x,n_y=0}^{N-1} \left( \mu_{n_x,n_y}^{f_1} f_{1,\boldsymbol{i}}^\dagger f_{1,\boldsymbol{i}} + \mu_{n_x,n_y}^{f_2} f_{2,\boldsymbol{i}}^\dagger f_{2,\boldsymbol{i}} \right) e^{2iq(n_x x + n_y y)} \right],\end{aligned} \tag{C1}$$

with $\boldsymbol{A}_{\boldsymbol{i}} = Hx(0,1,0)$, and $e_{\text{qsl}} = 0$ if $b=0$, and $e_{\text{qsl}} = 1$ otherwise. We self-consistently determine the parameters $\mu^c$, $\mu_{n_x,n_y}^{f_1}$, and $\mu_{n_x,n_y}^{f_2}$ for $H=0$. We exploit the residual translation symmetry $\boldsymbol{i} \to \boldsymbol{i} + N\hat{\boldsymbol{y}}$ to Fourier transform along the $y$-direction. We therefore cast Hamiltonian (C1) in the form

$$\mathcal{H} = \sum_{\sigma} \sum_{i_x,j_x} \sum_{k_y}{}' \sum_{\ell,\ell'=1}^{3N} \Psi_{i_x,k_y,\ell,\sigma}^\dagger \, \mathcal{H}_{i_x,j_x}^{\ell\ell'}(k_y, H) \, \Psi_{j_x,k_y,\ell',\sigma} \tag{C2}$$

where we have defined $\sum_{k_y}' = \int_0^{\frac{2\pi}{N}} \frac{dk_y}{2\pi}$, and the $3N$-component spinor

$$\Psi_{i_x,k_y,\sigma} = \begin{pmatrix} \widetilde{\Psi}_{i_x,k_y,\sigma}^0 & \dots & \widetilde{\Psi}_{i_x,k_y,\sigma}^{N-1} \end{pmatrix}^T, \tag{C3a}$$

$$\widetilde{\Psi}_{i_x,k_y,\sigma}^\ell = \begin{pmatrix} c_{i_x,k_y+2\ell q,\sigma} & f_{1,i_x,k_y+2\ell q,\sigma} & f_{2,i_x,k_y+(2\ell-1)q+\frac{\pi}{2},\sigma} \end{pmatrix}. \tag{C3b}$$

For $\eta \to 0^+$, the density of states at the Fermi level is defined as

$$\rho\left(H^{-1}\right) = -\frac{1}{\pi}\text{Im}\left\{ \sum_{k_y}{}' \text{Tr}\left[ \left(i\hat{I}\eta - \hat{\mathcal{H}}(k_y, H)\right)^{-1} \right] \right\}, \tag{C4}$$

where $\hat{\mathcal{H}}(k_y, H)$ is a matrix both in in the real space index $i_x$ and in the index $\ell$ of the spinor (C3), and $\hat{I}$ is the $(3NN_x) \times (3NN_x)$ identity matrix, with $N_x$ the number of lattice sites in the $x$-direction. Imposing open boundary conditions along $x$, and considering up to fourth-neighbor hoppings in $t^c_{ij}$ and $t^f_{ij}$, $\hat{\mathcal{H}}(k_y, H)$ is a block tridiagonal matrix, each block having linear dimension $2 \times 3N$. In total, there are $N_x/2$ of these blocks along the main diagonal. We exploit this structure to employ an iterative algorithm to compute Eq. (C4), as described in Refs. [35, 77]. The matrix in Eq. (C4) can be cast in the form

$$\hat{\mathcal{H}}(k_y, H) - i\eta\hat{I} = \begin{pmatrix} h_{11} & t_{12} & 0 & \ldots \\ t_{12}^\dagger & h_{22} & t_{23} & \ldots \\ 0 & t_{23}^\dagger & h_{33} & \ldots \\ \vdots & \vdots & \vdots & \end{pmatrix}, \tag{C5}$$

with

$$h_{nn} = \begin{pmatrix} \mathcal{H}^{\ell\ell'}_{2n,2n}(k_y, H) - i\eta\,\delta_{\ell\ell'} & \mathcal{H}^{\ell\ell'}_{2n,2n+1}(k_y, H) \\ \mathcal{H}^{\ell\ell'}_{2n+1,2n}(k_y, H) & \mathcal{H}^{\ell\ell'}_{2n+1,2n+1}(k_y, H) - i\eta\,\delta_{\ell\ell'} \end{pmatrix}, \tag{C6}$$

and

$$t_{n,n+1} = \begin{pmatrix} \mathcal{H}^{\ell\ell'}_{2n,2n+2}(k_y, H) & 0 \\ \mathcal{H}^{\ell\ell'}_{2n+1,2n+2}(k_y, H) & \mathcal{H}^{\ell\ell'}_{2n+1,2n+3}(k_y, H) \end{pmatrix}. \tag{C7}$$

We can efficiently calculate the density of states as

$$\rho(H^{-1}) = \frac{1}{\pi}\mathrm{Im}\left\{ {\sum_{k_y}}' \sum_{n=1}^{N_x/2} (h_{nn} - L_n - R_n)^{-1} \right\} \tag{C8}$$

where $L_n$ and $R_n$ are iteratively determined as

$$L_{n+1} = t_{n,n+1}^\dagger (h_{nn} - L_n)^{-1} t_{n,n+1}, \tag{C9a}$$
$$R_{n-1} = t_{n-1,n} (h_{nn} - R_n)^{-1} t_{n-1,n}^\dagger, \tag{C9b}$$

with the initial conditions $L_1 = 0$ and $R_{N_x/2} = 0$.

We discretize $k_y$ with $N_{k_y}$ points uniformly spaced in the interval $[0, 2\pi/N)$, corresponding to a system with $N_y = N \times N_{k_y}$ lattice sites in the $y$-direction with periodic boundary conditions. For the plots shown in the main text and here, we have fixed $\eta = 2.5 \times 10^{-4}$ eV, $N_{k_y} = 24$, and $N_x = 2 \times 10^3$.

### Appendix D: Plots of the Fermi surfaces in the reduced BZ

We have calculated the eigenenergies of Hamiltonian (B1) and the corresponding Fermi surfaces for every $\mathbf{k}$ in the reduced BZ. We have subsequently "unfolded" the Fermi surfaces to obtain Fig. 3 of the main text and Fig. 7. Fig. 5 illustrates how Fig. 3 of the main text has been obtained from the eigenenergies of Hamiltonian (B1): we first plot the eigenergies (gray lines) in the full BZ, and we then mark in red (or blue) the contours corresponding to the $\alpha$, $\beta$, and $\delta$ pockets, sometimes crossing the borders between neighboring reduced BZs. In Fig. 6, we show the Fermi surfaces in the reduced BZs, together with their orbital content. We observe that the $\alpha$ pocket is essentially made by the $c$-electrons, while the $\beta$ and $\delta$ pockets have their most weight on the $f_1$ fermions.

### Appendix E: Additional results for different choices of the parameters

In this section, we show plots of the Fermi surfaces for different choices of the parameters $b$ and $V$, compared to the main text. In Fig. 7, we show that the structure of the Fermi surfaces shown in Fig. 3 of the main paper is not qualitatively affected by different choices of the parameters, both in the CDW* (panel a) and in the CDW (panel b) phases. In the former, we recognize the $\alpha$ and $\beta$ electron pockets, whereas in the latter we have an $\alpha$ electron pocket and two $\delta$ hole pockets, as well as small pockets centered around the locations where the nodal points of the $\pi$-flux spin liquid would emerge in absence of CDW order.

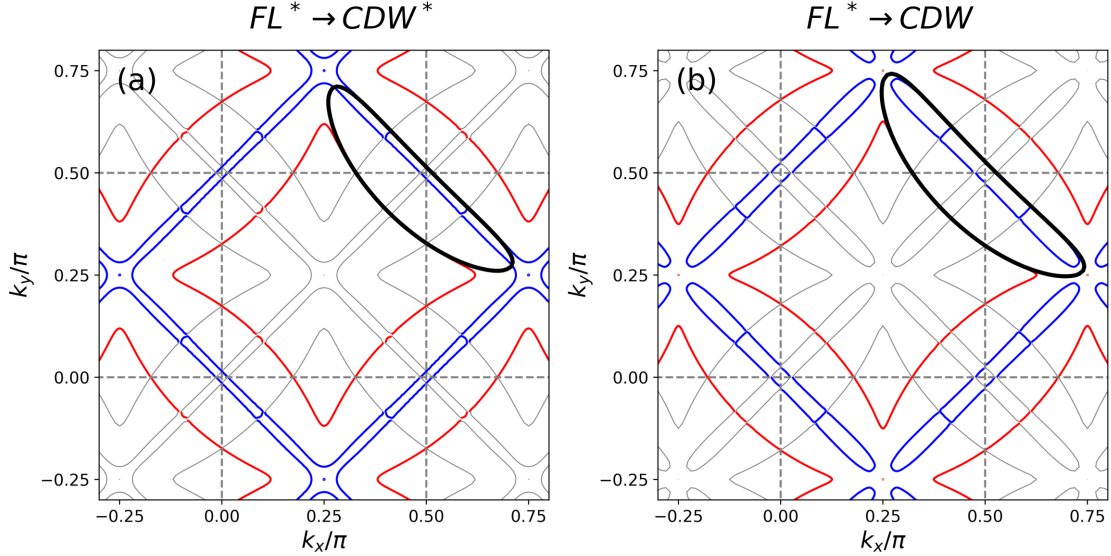

FIG. 5: Same as in Fig. 3(b) (panel a) and Fig. 3(c) (panel b) of the main text. In addition, we show (gray lines) the zeros of the eigenenergies of Hamiltonian (B1).

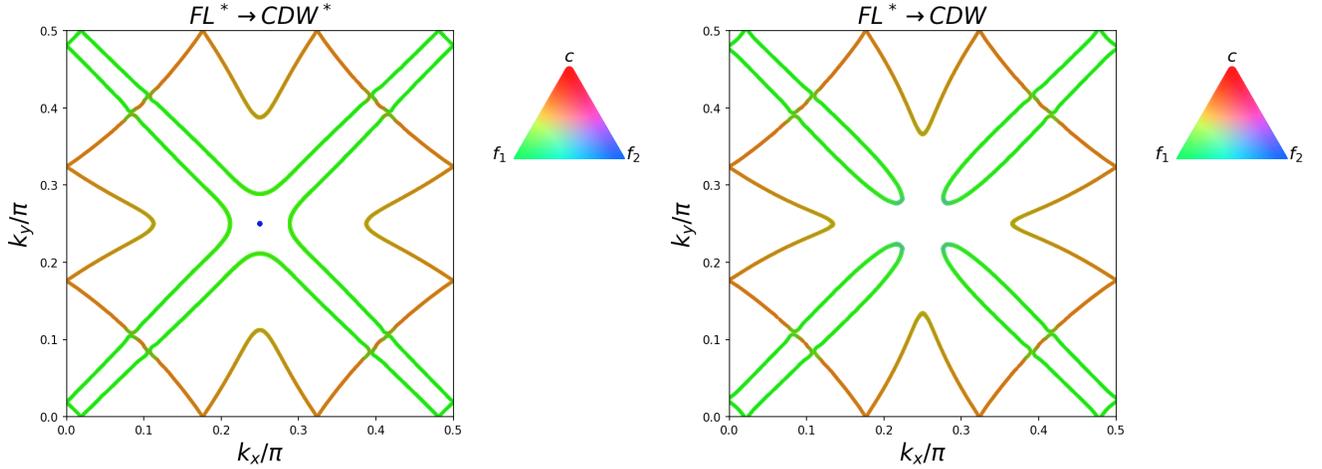

FIG. 6: Fermi surfaces in the reduced BZ $k_x, k_y \in [0, \pi/2)$ for the CDW* phase (left panel) and the CDW phase (right panel). The parameters are the same as in Figs. 3 and 4 of the main paper. The color codes (see insets) indicate the orbital weight of the $c$, $f_1$ and $f_2$ fermions on the Fermi surfaces.

In Fig. 8, we show the quantum oscillations (insets) of the density of states $\rho$ as a function of the inverse magnetic field, and its Fourier spectrum (main plots). In panel (a), we observe that a stronger modulated potential in the CDW* phase further enhances the peak at $f_\beta$, compared to what we have shown in Fig. 4a of the main paper, but the qualitative structure remains unaffected. In panel (b), with parameters corresponding to a CDW phase, we observe that, compared with Fig. 4b of the main text, a stronger CDW order parameter suppresses the magnetic breakdown peak at $f = f_\gamma$, while enhancing the peak at $f_\delta$, which could be anyway beyond the experimental resolution, due to its low frequency $f_\delta \simeq 0.002 f_{\rm BZ} \simeq 55$ T.

### Appendix F: Additional results for a period-6 bi-directional CDW

In this section, we present results for period-6 CDW* and CDW states. The formalism is the same as in Secs. A and B, except that we consider $q = \frac{\pi}{6}$, and $N = 6$. We fix the density to $n = 0.945$, such that the hole pockets have

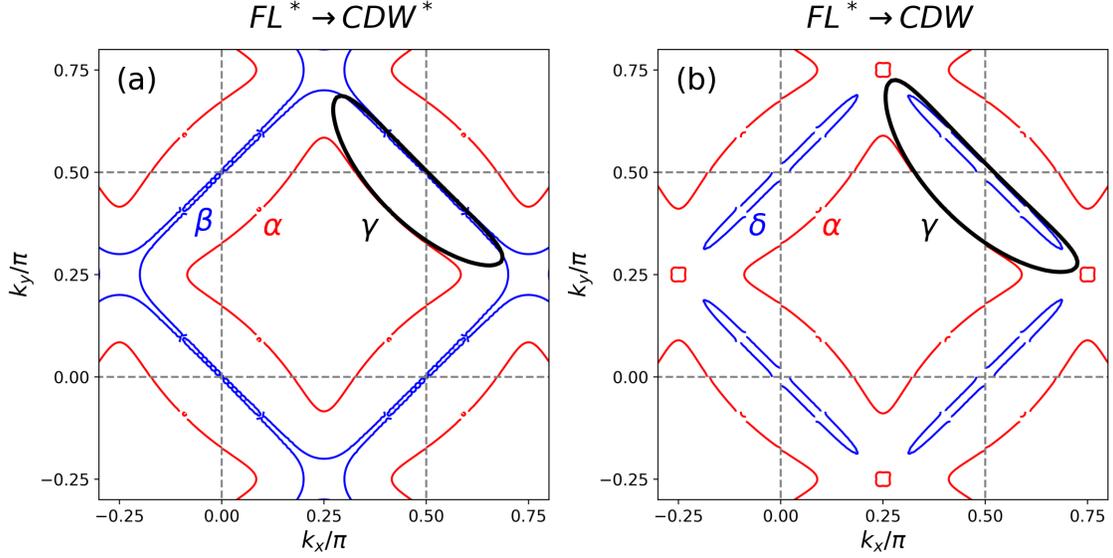

FIG. 7: Same as in Fig. 3(b)-(c) of the main paper but with the parameters $V = 0.04$ eV, $b = 0$ in panel (a), and $b = 0.2$ eV, $g = 1$, $V = 0.03$ eV in panel (b).

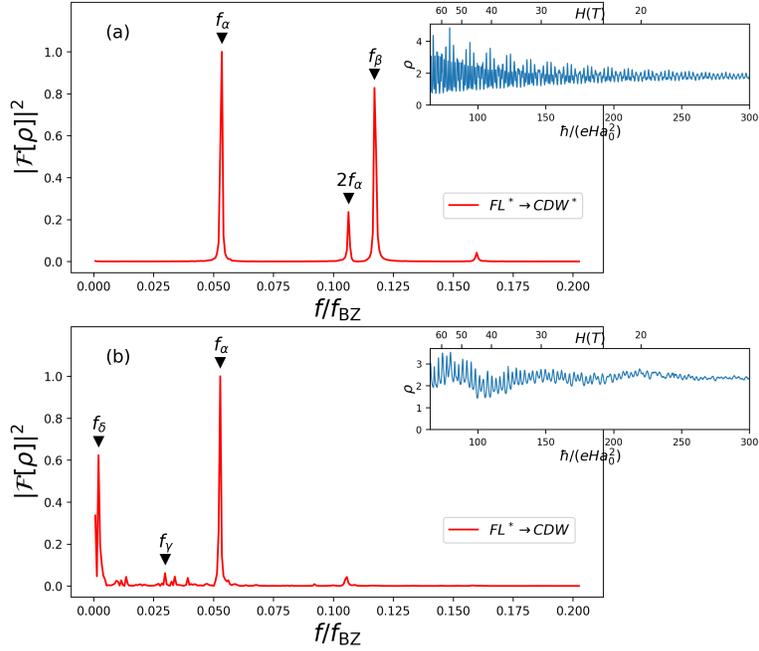

FIG. 8: Quantum oscillations of the density of states (insets) and their Fourier spectra (main plots). The parameters in panel (a) and (b) are the same as in Fig. 7(a) and Fig. 7(b), respectively.

nesting vectors $(2\pi/3, 0)$ and $(0, 2\pi/3)$. For this reason, differently than for the $N = 4$ case, we define the potential in the top layer as

$$V \sum_{\boldsymbol{i},\sigma} [\cos(4qx) + \cos(4qy)] \, c^\dagger_{\boldsymbol{i},\sigma} c_{\boldsymbol{i},\sigma} \,. \tag{F1}$$

Since for $V = 0$ and $b = 0$, Hamiltonian (B1) has nodal points of the $f_2$ spinons at $(n_x \frac{\pi}{3}, n_y \frac{\pi}{3})$ (instead of $((2n_x + 1)\frac{\pi}{4}, (2n_y + 1)\frac{\pi}{4})$ for the $N = 4$ case), in order to have a nodal point at the center of the reduced BZ,

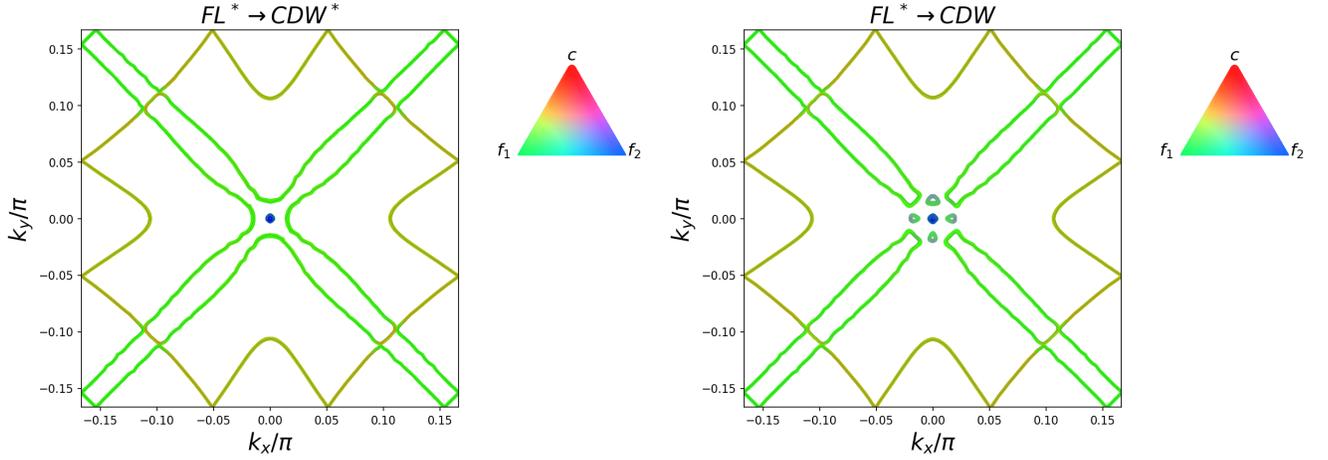

FIG. 9: Fermi surfaces for a period-6 CDW* state (left panel), and CDW state (right panel). The parameters are $b = 0$, $V = 0.02$ eV in the left panel, and $b = 0.1$ eV, $g = 0.5$, $V = 0.02$ eV in the right one. The color codes (see insets) indicate the orbital content of the Fermi surfaces.

we define it as $(k_x, k_y) \in (-\pi/6, \pi/6]$. In Fig. 9, we show the Fermi surfaces in the reduced BZ for a period-6 CDW* and CDW. Comparing Fig. 9 with Fig. 6, we see that the qualitative structure observed remains the same. In the CDW* state, we find an $\alpha$ pocket and a larger $\beta$ pocket, while in the CDW phase we observe the $\alpha$ pocket, two small elongated $\delta$ pockets as well as five extremely small pockets around $(k_x, k_y) = (0, 0)$ with a significant contribution of the $f_2$ fermions.

In Fig. 10, we show the quantum oscillations (insets) and the power spectrum (main plots) for a period-6 CDW* phase (panel a) and for a period-6 CDW phase (panel b). The parameters we used are the same as in Fig. 9. In both cases we observe a dominant peak at the frequency $f_\alpha \simeq 0.024 f_{\text{BZ}}$, corresponding to the $\alpha$ pocket. Note that, compared to the results presented in Fig. 4, this frequency, corresponding to about 650 T, is much closer to the values 530 T [26, 31], 660 T [25], and 840 T [29] observed in YBa$_2$C$_3$O$_{6+\delta}$ at hole doping $p = 0.11$, YBa$_2$Cu$_4$O$_8$ at $p = 0.125$, and HgBa$_2$CuO$_{4+\delta}$ at $p = 0.09$, respectively. The mismatch between the experimentally observed frequency and our theoretical calculation has two main reasons. First of all, we have used hopping parameters that best reproduce another compound, namely Bi2201. Moreover, we are working at a different doping level, namely $p = 0.055$ here and $p = 0.155$ in the main text. In the CDW* state, Fig. 10a, we also observe a pronounced peak, whose height is about 30% of the height of the peak at $f_\alpha$, at $f_\beta \simeq 0.05 f_{\text{BZ}}$, corresponding to the $\beta$ pocket, as well as higher harmonics of $f_\alpha$. In the CDW phase, Fig. 10b, a magnetic breakdown frequency at $f_\gamma \simeq 0.01 f_{\text{BZ}}$ appears, with a peak height of about 14% of the peak at $f_\alpha$. This frequency roughly corresponds to the volume of the hole pockets in absence of CDW order. A tiny peak (3% of the peak height at $f_\alpha$) appears at $f_\beta$, probably due to magnetic breakdown processes. A low peak at $2f_\alpha$ is also present.

### Appendix G: Charge and bond density profiles

In this section, we show how the charge and bond density profiles appear in the period 4 and period 6 CDW phases, where the spinons are confined. The charge and bond densities are defined as

$$\varrho_{\bm{i}}^c = \sum_{n_x,n_y=0}^{N-1} \sum_{\mathbf{k}} \langle c_{\mathbf{k}}^\dagger c_{\mathbf{k}+2q(n_x,n_y)} \rangle \, e^{2iq(n_x \cdot x + n_y \cdot y)} \,, \tag{G1a}$$

$$\varrho_{\bm{i}}^{b,x} = \sum_{n_x,n_y=0}^{N-1} \sum_{\mathbf{k}} \cos k_x \langle c_{\mathbf{k}}^\dagger c_{\mathbf{k}+2q(n_x,n_y)} \rangle \, e^{2iq\left(n_x \cdot \left(x+\frac{1}{2}\right) + n_y \cdot y\right)} \,, \tag{G1b}$$

$$\varrho_{\bm{i}}^{b,y} = \sum_{n_x,n_y=0}^{N-1} \sum_{\mathbf{k}} \cos k_y \langle c_{\mathbf{k}}^\dagger c_{\mathbf{k}+2q(n_x,n_y)} \rangle \, e^{2iq\left(n_x \cdot x + n_y \cdot \left(y+\frac{1}{2}\right)\right)} \,, \tag{G1c}$$

where $\varrho_{\bm{i}}^c$, $\varrho_{\bm{i}}^{b,x}$, and $\varrho_{\bm{i}}^{b,y}$ are the charge, $x$-bond and $y$-bond densitites, respectively. The expectation value $\langle \ldots \rangle$ are taken with respect to the mean-field Hamiltonian (B1). In Fig. 11, we show the charge and bond density profiles

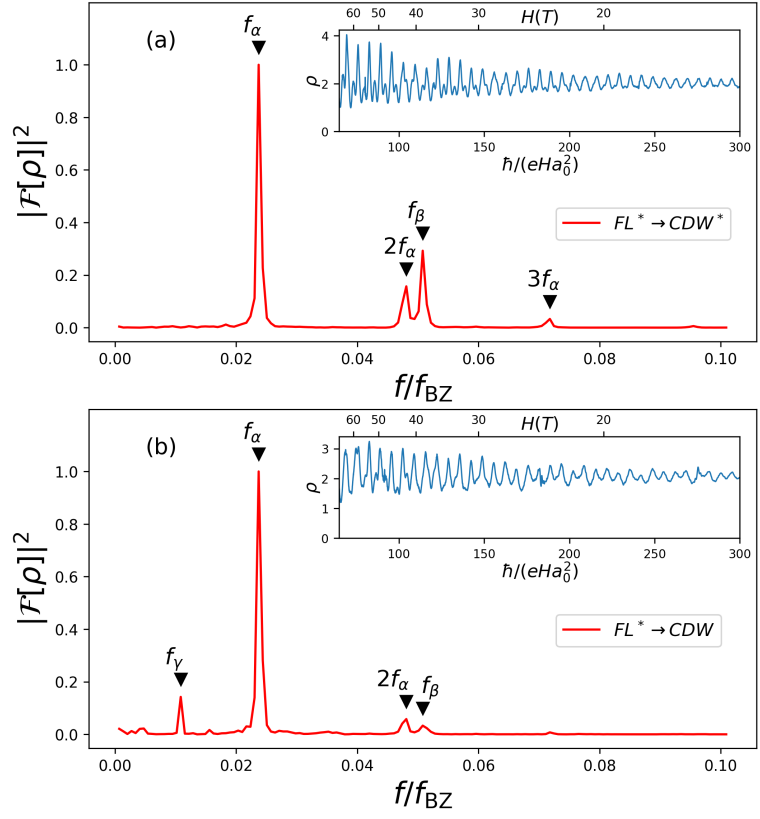

FIG. 10: Quantum oscillations of the density of states (insets) and their Fourier spectra (main plots) for a period-6 CDW* phase (panel a) and a period-6 CDW phase (panel b). The parameters of panel (a) and (b) are the same as in Fig. 9a and 9b, respectively. The density of states has been computed on a finite size system with $1333 \times 96$ sites.

for a period 4 (left panel) and period 6 (right panel) CDW phase. We see that in the period 6 CDW, even though the chargon condensate has a period 6, the modulated potential (F1), having period 3, couples more strongly to the electrons, giving the impression of a period 3 CDW.

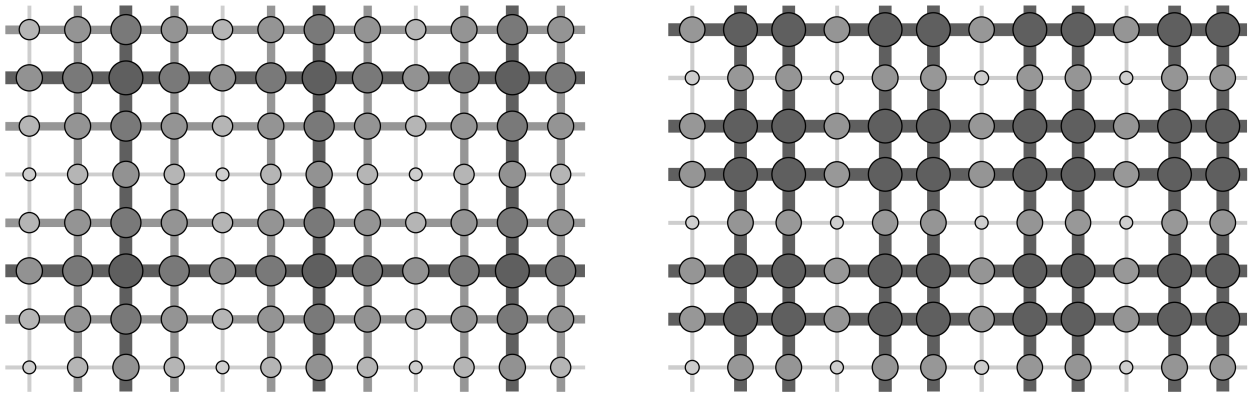

FIG. 11: Typical charge and bond density profiles in a period 4 CDW state (left) and in a period 6 CDW state (right). Larger and darker dots and thicker and darker lines represent higher charge and bond densities, respectively, whereas smaller and brighter dots and thinner and brighter lines represent lower charge and bond densities, respectively. Note that the period 6 state appears to have period 3 because the period 3 potential $V \sum_i (\cos(2\pi x/3) + \cos(2\pi y/3)) c_{\bm{i}}^\dagger c_{\bm{i}}$ couples more strongly to the electron charge and bond density than the chargon condensate. The parameters are $b = 0.2$ eV, $g = 1$, $V = 0.03$ eV for the left panel, and $b = 0.1$, $g = 0.5$, and $V = 0.02$ eV for the right panel.



\end{document}